\documentclass[sigconf]{aamas} 

\usepackage{balance} 
\usepackage{algorithm2e}
\RestyleAlgo{ruled}

\setcopyright{ifaamas}
\acmConference[AAMAS '23]{Proc.\@ of the 22nd International Conference
on Autonomous Agents and Multiagent Systems (AAMAS 2023)}{May 29 -- June 2, 2023}
{London, United Kingdom}{A.~Ricci, W.~Yeoh, N.~Agmon, B.~An (eds.)}
\copyrightyear{2023}
\acmYear{2023}
\acmDOI{}
\acmPrice{}
\acmISBN{}
\subtitle{Extended Abstract}

\acmSubmissionID{1148}

\title{A Theory of Mind Approach as Test-Time Mitigation Against Emergent Adversarial Communication}

 \author{Nancirose Piazza}
 \affiliation{
   \institution{SAIL Lab - University of New Haven}
   \city{West Haven, CT}
      \country{United States}
   }
 \email{npiazza@newhaven.edu}

 \author{Vahid Behzadan}
 \affiliation{
   \institution{SAIL Lab - University of New Haven}
   \city{West Haven, CT}
      \country{United States}
   }
 \email{vbehzadan@newhaven.edu}

\begin{abstract}
Multi-Agent Systems (MAS) is the study of multi-agent interactions in a shared environment. Communication for cooperation is a fundamental construct for sharing information in partially observable environments. Cooperative Multi-Agent Reinforcement Learning (CoMARL) is a learning framework where we learn agent policies either with cooperative mechanisms or policies that exhibit cooperative behavior. Explicitly, there are works on learning to communicate messages from CoMARL agents; however, non-cooperative agents, when capable of access a cooperative team's communication channel, have been shown to learn adversarial communication messages, sabotaging the cooperative team's performance particularly when objectives depend on finite resources. To address this issue, we propose a technique which leverages local formulations of Theory-of-Mind (ToM) to distinguish exhibited cooperative behavior from non-cooperative behavior before accepting messages from any agent. We demonstrate the efficacy and feasibility of the proposed technique in empirical evaluations in a centralized training, decentralized execution (CTDE) CoMARL benchmark. Furthermore, while we propose our explicit ToM defense for test-time, we emphasize that ToM is a construct for designing a cognitive defense rather than be the objective of the defense. 
\end{abstract}


\newcommand{\BibTeX}{\rm B\kern-.05em{\sc i\kern-.025em b}\kern-.08em\TeX}

\begin{document}


\pagestyle{fancy}
\fancyhead{}


\maketitle 


\section{Introduction}
Multi-agent systems can be found in many prominent domains that is essential to the functionality of society, extending recent deployment of automation to autonomous mobile vehicles\cite{zhou2020smarts}, intelligent transportation systems\cite{10.1016/j.eswa.2014.05.015}, financial trading portfolios\cite{Lee_2020}, and joint user scheduling in cloud-edge systems for the minimization of latency and instaneous Quality of Service (QoS) for user equipment\cite{9340599}. Multi-agent approaches to such problems enables the transformation of these complex tasks and computationally infeasible into distributive, paralleled and locally manageable workloads. In tasks that involve the learning of sequential strategies and behaviors, promising approaches have risen from the framework of Multi-Agent Reinforcement Learning (MARL). A comprehensive survey on MARL by \cite{s1} supports the long lived interest and diverse perspectives on the approaches towards MARL, eg. value-decomposition, policy-gradients and communication. A more recent paper by \cite{https://doi.org/10.48550/arxiv.1911.10635} provides a selection of theories and algorithms relevant to current research.

The study of agent-to-agent relationships is often described as cooperative or non-cooperative (eg. competitive). Agents that are fully cooperative, often modeled as team games, can have shared rewards. Some value-based works for fully-cooperative MARL tasks (eg. \cite{agogino2004}, \cite{son2019},\cite{sunehag2017}, \cite{rashid2018}, \cite{mguni2019}) consider estimating individual value functions of agents, global value functions for the shared reward, and/or reward shaping through some mechanism. On the other end, agents can be described as fully competitive where the maximization of an agent's local reward results in the payoff loss of another agent's reward. Between fully competitive and fully cooperative is mixed where agents can demonstrate both competitive and cooperative behaviors.

Communication, as outlined as a pillar for cooperation intelligence, is an important channel allowing for negotiation (eg. emergence of communication through negotiation \cite{cao2018}) and sharing or transferring information. CoMARL often uses communication in the form of message transferring to address partially-observable information and/or signaling cues for coordination. Some design architectures for CoMARL in-cooperate graph models to aggregate various inputs for decision-making, often following the CTDE paradigm. However, such mechanisms and design can be exploited,Work by \cite{blumenkamp2020emergence} showed that self-interested agents can learn to communicate adversarial, messages to a cooperative team's communication channel, impairing the cooperative team's performance. Extending to real world scenarios, successful adversarial information transfer can result in cascading failure and catastrophic events (eg. impact of cascading failure in complex networks \cite{10.1093/comnet/cnaa013}.)

The contributions of this paper is as followed:

\begin{itemize}
    \item We present Theory of Mind (ToM) as a cognitive mechanism for defense against adversarial communication, explicitly our ToM design is built from historic observable neighbor actions.
    \item We present an belief-based trust defense test-time mechanism for homogeneous-policy agents or agents that have secured access to other cooperative agents' policies. Such defense requiring no additional training and will allow agents to form decentralized ToM trust beliefs over other agents based on observable neighboring actions to choose which agents' communicated message will be leveraged.

    \item We present empirical results from an adversarial communication environment, CoverageEnv.
    \item We discuss where this implementation fits within current taxonomy and comparison to other similar literature, distinguishing the differences.
\end{itemize}

\section{Problem Statement}
We define a MARL setting by a tuple:
$$<V,A,S,P,\{R^{i}\}_{i \in V},\{O^{i}\}_{i \in V}, O, \gamma>$$
where $V$ is the set $\{1,2...|V|\}$ of agents, $A$ is the action space, $S$ is the state space, $P$ is the transition probability matrix, $R^{i}:(i_a)\rightarrow \mathbb{R}$ is the local reward function of agent $i \in V$ taking action $a$, $O^{i}$ is agent $i$'s transmitted observation in a state $s$, $O$ is  agent $i$'s ground-truth observation in a state $s$, and $\gamma \in [0,1)$ is a discount factor.

Furthermore, we search for a joint policy $\pi: (\pi_1,\pi_2 ... \pi_{|V|})$ which produces joint action $(a_1, a_2 , ..., a_{|V|})$ which maximizes the local reward for each cooperative, homogeneous agent.

A graph network of agents $G = <V,E>$ where $V$ is the set of agents $\{ 1,2...|V|\}$ and $E$ is the set of existing edges between any pair of communicating agents $V \times V $, possibly dynamically changing over time. The edge set $E$ would be the set of bidirectional decentralized communication channels between pairs of agents $(i,j)$ where agent $i$ transmits $O^{i}$(agent $i$'s observation) to agent $j$. This observation or transmitted message can be any information such as the local observation/state, engineered information like trust scores, local reward, etc. In decentralized execution, encoded information is shared in communication instead of local observation but we include local observation in communication such that our agents can self-evaluate others' observations. In other words, we shift from letting an agent directly trust other agents' encoded information and instead require the transmission of state observation which is understandable by the designer. Further traditional, non-machine learned filtering defenses can be used on the state observation to minimize convincing adversarial messages by allowing the designer to understand valid perturbations.

Our approach to mitigating against adversarial communication is through a history of trust mechanism, updating its probability of trust in a direction and magnitude related to its previously witnessed self-committing actions of other agents with emphasis on Theory of Mind as its objective. Similar to human reasoning, agents reason of other agents' observed behavior and update their belief based on past evidence, for example, credit history. For our work, we consider homogeneous agents (according to the environment), but this can be extended to any team of heterogenous agents given their evaluators are available and public to its teammates. An analogy is where a person's public encryption key for email may be uploaded to a shared keyserver for others.

\section{Related Work}

Learning to communicate in MARL settings\cite{foerster2016learning} reduce difficult tasks possibly due to partial observation into a more complete information set, maximizing the success of task competition. However, the attack surface enlarges when a communication channel can directly impact a subset of agents, emphasizing the importance of awareness for adversarial communication. Furthermore, because these agents are semi-supervised, without proper constraints to avoid learning bad behaviors, behaviors such as adversarial communication in MARL\cite{blumenkamp2020emergence} can emerge and is difficult to address without a priori.

Modeling trustworthiness is a common direction for defenses. For example, a paper called modeling context aware dynamic trust using hidden markov models by \cite{10.5555/2900929.2901002} which models agent interactions and ideally agent intent. Our work also fits under modeling trustworthiness, but we take a different direction towards choosing models which are inspired by credibility. One relevant work on leveraging observed past actions of other agents as an initial trust model is from Schillo et al \cite{doi:10.1080/08839510050127579}. In their paper, they also leverage the observed actions of other agents and while a consensus-based defense is possible (we have considered it), we leave it in favor for a more self-evaluation type of trustworthiness model to minimize the need for external input.

A recent survey by \cite{prorok2021robustness} towards a standardization in taxonomy and establishment for robustness in MARL agents present open problems for MARL and consolidates the existing work under their taxonomy. We believe our work would fit within the backend perception and filtering methods. The survey classifies approaches as pre-operative, intra-operative, and post-operative with respect to a stressor (disturbance eg. noise). Our approach updates beliefs during execution, implying it most likely would be an intra-operative approach with an out-of-distribution stressor; however our defense attempts to also address in-distribution stressor given the adversary is still interested in its own performance in the environment.

Defense by \cite{mitchell2020gaussian} for adversarial communication proposes a variation auto-encoder bayes model as a message filtering method for weighing messages of other agents based on confidence on truthfulness derived from mutual information and consistency of information. Their work is close to how we approach the problem through probabilistic beliefs; however, instead of crafting a belief state and require training of another model, our belief state is dependent on observable behavior during episodic execution. Another work by \cite{10.5555/3463952.3463996} uses consensus-based decisions to detect and eliminate adversarial robots in a flocking problem. While their approach is more reactant to adversarial detection and dependent on an uncompromised majority, their solution is a centralized method of evaluating trust, contrast and complimentary to our decentralized method.

Alternatively, complete adversarial agents may not be realistic. Instead, it may be more feasible for a cooperative agent to be hijacked, attacked for a segment of time, or beyond intentional malice such as uncertainty. In this case, work by \cite{Zhou_2021} lays out current directions in risk-aware agents and state resilience when facing uncertainty. Many approaches for uncertainty can be leveraged for resilience to non-cooperative behavior, including adversarial. For example, an approach for a more resilient flock of mobile robot agents through resilient formation control from a distributed resilient controller \cite{Saulnier2017ResilientFF} in face of non-cooperative or defective agents. Detecting adversarial agents can also take more of an authentication approach. Work by \cite{8250942} use physical fingerprints and the analysis of these fingerprints from legitimate neighboring agents to identify spoofing adversarial agents.


\section{Self-Revealing Behaviors }
Suppose we have a set of two-player games with a known distribution across games but the player valuation functions are private. Players can perform `cheap talk' where they can verbally communicate a message, in this case, a message contains a statment on the action they will take. Furthermore, a message used for communication can be described as \textit{self-revealing} or \textit{self-committing} when the action stated in the message can be observed. A message from $p_1$ that is self-revealing informs $p_2$ about the current game $G_i$. With immediate payoff, $p_2$'s message reveals to $p_1$ whether their relationships may be cooperative or competitive, implying information about the current game. A message from $p_1$ is self-committing when it is consistent with $p_2$'s observation of $p_1$'s action $a_1$. This commitment is important to agent conflict resolution areas of research such as negotiations\cite{10.1017/S0269888999003021}, though for our work, we do not attempt resolutions of conflict, but rather predict agent roles based on behaviors. To keep our work's vocabulary aligned with other works, we will say an action is \textit{consistent} if the following observed action matches the expected action derived from the previous message.
By describing messages as consistent, we can build a trust history of other agents in a similar manner humans do. 

While learning to communicate is ideally used to encourage cooperative learning behavior, signals do not have to be self-committing, in other words, one can either be babbling where no information can be deduced from the message or deceitful as demonstrated with \cite{blumenkamp2020emergence}. While one does not expect non-committing signals in cooperative tasks; in the presence of misaligned, trained policies trained in cooperative settings may be maladjusted and brittle in handling non-cooperative behavior. Furthermore, miscommunication may also occur due to noise in transmissions which should be accounted for resilience. One method which can help team agents behave in the presence of non-cooperative behavior is to create a belief over their histories which is composed of both messages from others and not directly observed actions of other agent's actions. Similar to \cite{mitchell2020gaussian} where some empirical belief determines trustworthiness, we consider that if other agents' messages are not self-committing, then communication messages should not impact an agent's decision.

\section{Theory of Mind Under Test-Time}
Theory of Mind (ToM) is the rationalization of other agents to some belief state, often through Bayesian modeling. Bayesian models have belief states that are used to approximate probabilistic frequencies for sequences of outcomes given the discrete or continuous measure of time and some public information. A well-known Bayesian estimation algorithm in control theory is the Kalman filter, see work by \cite{7528889} for a survey on the applications, which estimates unknown variables and uncertainties over a time sequence of observations. While we do not use the Kalman filter, it can certainly be used as a ToM defense for systems with target tracking such as autonomous motor vehicles. Furthermore, the extension to Kalman filter may be more difficult to model when agents have learned encoding messages. ToM is well discussed in AI; however, explicitly modeling other agents' belief states can require additional burdensome computational load that may not be feasible in large-scaled systems of agents such as swarm robotics\cite{Schranz2020SwarmRB} where ideally each agent is otherwise not computationally expensive. However, Mean-Field MARL (MF-MARL)\cite{yang2018} can be a substituting framework for large $N$ agents at which we only record observable actions from agents within a defined radius.

We emphasize that ToM should be influential to a defense's design rather than declare one explicit interpretation of ToM, namely the direct modeling of other agent's belief state, as the sole representation of how it should be implemented.

We approach ToM from the perspective of cooperative intelligence where an agent follows the reasoning `\emph{if I were them: what would I do, what message would I transmit to cooperate, and whether their observed behavior is consistent with what I expect given their transmitted message?}'

Ideally, if agent $i$ observes agent $j$'s behavior to be consistent with their own expectation, then agent $i$ may be inclined to trust agent $j$'s communication message and leverage it. In the case that it is inconsistent, the message can be discarded.

Many defenses rely on out-of-distribution detection, including the message filtering VAEB defense \cite{mitchell2020gaussian}. However, these type of defenses require additional training outside of test-time, partially because it focuses on transforming high-dimensional data. Instead, we direct our attention to observable agent actions which is of low dimension. Our ToM belief defense, similar to human credit history, use the history of observed actions to update the probability of being trusted. This defense is directed towards test-time because an agent's belief is only relevant to the current episode and episode's history, in contrast to the traditional approach of learning over episodes. Furthermore, we believe defenses should be layered to minimize possible damage from adversarial communication, implying our defense is complimentary to traditional defenses.

\section{Proximal Policy Optimization}
There are two methods of deriving a policy both in multi-agent and single-agent RL: value-based and policy-based. Value-based methods optimize parameters to estimate an optimal $Q^{*}$ function, in multi-agent settings this may be a joint $Q$ function which is used to extract a joint policy $\pi$. Policy-gradient methods optimize parameters through gradient descent to maximize the expected log-likelihood over actions, directly $\pi$. Actor-critic is a method that optimizes an actor/policy $\pi$ with assistance from a critic/value-based function ($Q$,$V$, or $A$/Advantage). A cooperative policy gradient is formulated as:

$$ g^{i}_{k} = \mathbb{E}_{\pi} \bigg[ \underset{j \in V}{\sum} \nabla_\theta \pi_{\theta}^{j}(a^{j} | O)A^{i}(s) \bigg] $$

which has been proven to converge to a local maximum of the expected sum of return for all agents with probability one\cite{blumenkamp2020emergence} and the joint policy maximizes the sum of cumulative rewards.

\section{Public Valuation for Consistent Actions}
Given the assumption of an actor-critic architecture, we assume each agent have access to a value function $Q$, $V$ or $A$. Though in practice, deployment of agents entail just the actor/policy itself, we allow our agents to also include the value function $\hat{V}$. We define the $\hat{\rho}$ difference in value as:
\begin{equation}
    \hat{\rho} = \hat{V}(s_j,\pi_i(s_j)) - \hat{V}(s_j,\pi_j(s_j))
\end{equation}

where $\pi_j(s_j)$ is agent $i$'s observation of agent $j$'s action given $s_j$. Since the policies of agent $i$ and $j$ are assumed to be homogeneous, it is valid to evaluate under agent $i$'s policy. We introduce a threshold $\rho$ where if evaluated for the message $m_{ij}^{t}$ at timestep $t+1$ where $\hat{\rho} > \rho$, agent $j$'s action is considered not consistent. Likewise, if it falls within the threshold, the action is considered consistent.

The extension to stochastic policies can use an empirical collection of samples to estimate ${V}(s_j,\pi_i(s_j))$ during training such that we may create a distribution of calculated Kullback-Lieber measures for samples and use its mean as a threshold:

\begin{equation}
    KL(\hat{V}(s_j,\pi_j(s_j)||\hat{V}(s_j,\pi_i(s_j)))
\end{equation}


\begin{algorithm}[]
\SetAlgoLined

\KwResult{Update consistency count}
\For{agent $\in$ All$\_$Agents}{agent.trust = $\{0:0,1:0\}* $  len(All$\_$Agents) }
$ A_{c} \leftarrow $ \textit{critic's advantage function} \\
$a \leftarrow $ \textit{observed action} \\
$o^\prime \leftarrow $ \textit{ target agent communication}\\
evaluated$\_$idx $\leftarrow$ \textit{target agent idx} \\

 \For{cooperative$\_$agent $\in$ All$\_$Agents}{
 evaluator $\leftarrow$ cooperative$\_$agent\\
  \eIf{$\|\underset{a^\prime}{\max}A_{c}(o^\prime) - A_{c}(o^\prime,a)| <= \rho$}{
   evaluator.trust[evaluated$\_$idx][1] +=1
   }{
   evaluator.trust[evaluated$\_$idx][0] +=1
  }
 } 
 \caption{update$\_$consistency$\_$count}
 \label{algo_trust}
\end{algorithm}

\begin{algorithm}[]
\caption{update$\_$belief}\label{algo_consensus}
\SetAlgoLined
\KwResult{Adjust an agent's belief of another agent using the direction of previous observed action and the magnitude proportional to consistency count class.}

\For{agent $\in$ Coop$\_$Agents}{ 
\For{agent$\_$idx $\in$ All$\_$Agents}{
agent.belief[agent$\_$idx] = 1.0 // probability 1.0}}

$s \leftarrow$ \textit{learning rate}\\
\For{t $\in$ \{ 2,...N \} }{
$j \rightarrow$ \textit{interrogator agent} \\
$i \rightarrow$ \textit{interrogated agent} \\

\eIf{$\pi^j(o^i_{t-1}) = a^i_{t-1}$}{
    j.belief[i] + s $\times$ (j.trust[i][1] / t)  \textit{increase trust prob.}\\
    }
    {j.belief[i] - s $\times$ (j.trust[i][0] / t)  \textit{decrease trust prob.} }}
\end{algorithm}

The ideal implementation, outlined in Algorithm \ref{algo_trust} and Algorithm \ref{algo_consensus}, is as followed: each agent will hold a probability belief of all other agents, each agent will choose to trust other agents based on their individualized probabilities. All agents are set to probability 1.0 by default since there is no evidence for mistrust. After each environment step, all agents reevaluate their trust in other agents depending on the observed action from the environment step and adjust their belief based on whether the previous action was consistent and the magnitude is proportional to the running sum of observed consistent actions or non-consistent actions.

\section{Experimental Results}
We use the adversarial communication repository provided by \cite{blumenkamp2020emergence}, training under the default settings such as PPO actor-critic; however, we set the policy to always choose the action with the highest probability in the distribution. The coverage task is an environment where a team of agents are tasked to cover as much as the grid-world map as possible. The agents have access to a central communication channel where their extracted feature map is weighted by the graph shift operator (GSO) which determines the weight of neighboring nodes with an Aggregation Graph Neural Network (AGNN). In decentralized execution and decentralized communication, the GSO is calculated locally where we assume the validity of agent location can be verified by a third-party similar to a registration authority (RA).

We trained the cooperative policy using $N = 5$ agents for 6 million timesteps. The self-interested policy was trained for 6 million timesteps given the fixed cooperative policy. The re-adaption training continued with the fixed self-interested policy for another 6 million timesteps. The evaluation setup has one self-interested agent (Agent0) and three cooperative agents (Agent1, Agent2, Agent3).

\begin{figure}
    \centering
    \includegraphics[width=1.0\linewidth]{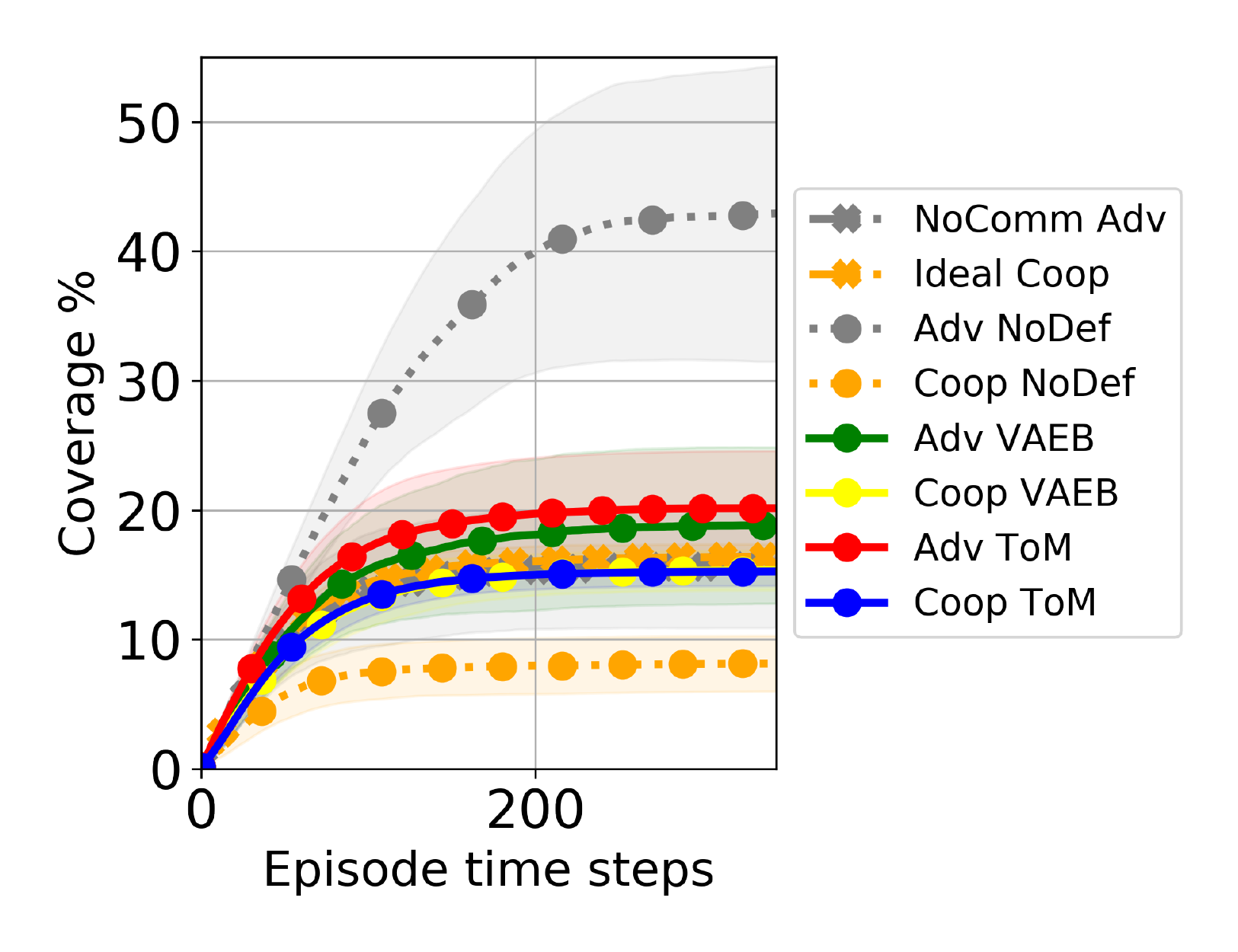}
    \caption{CoverageEnv - ToM performance comparison against an ideal defense baseline VAEB and no defense}
    \label{coverage1}
\end{figure}

\begin{figure}
    \centering
    \includegraphics[width=1.0\linewidth]{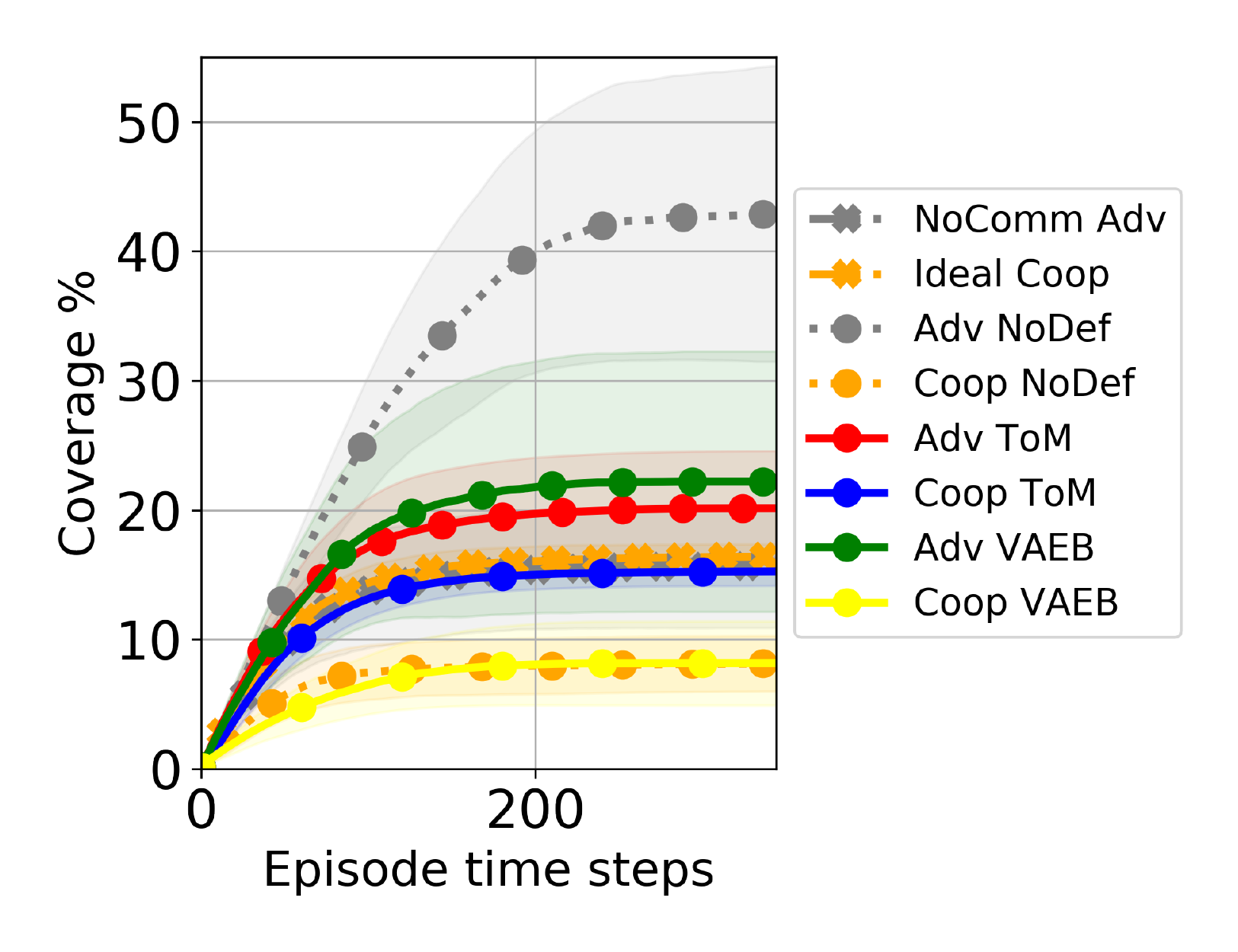}
    \caption{CoverageEnv - ToM performance comparison against a defense baseline VAEB after readaption training}
    \label{coverage2}
\end{figure}

We present Figure \ref{coverage1} as a baseline comparison for the \textit{ideal} performance of a variation auto-encoder bayes message filter (VAEB) trained on extracted samples/individual agent encoded feature maps from the training environment against our ToM defense (ToM) with a chosen learning step multiplier of 3.7, the secondary baseline of no defense (NoDef) and the ideal performance for the cooperative agents (Ideal Coop). Using generated sample $s$, we set that if  $||s||_2 > c$ for some chosen $c$ constant, to reject the message and otherwise accept the message. With ideal performance of the VAEB, ToM is comparable by coverage percentage performance in Figure \ref{coverage1}. Then, we show in Figure \ref{coverage2}, the VAEB is ineffective when implemented on top of cooperative agents that have re-adapted their communication messages in the presence of self-interested messages. This suggests that the newly learned communication most likely falls out of distribution of the VAEB, resulting in less truthful and useful information influencing agent decisions. This is exemplary that message filter methods like VAEB and the more complex Gaussian filter VAEB by \cite{mitchell2020gaussian} are rigid against non-stationary learned representations and will require additional training for distribution shifts. Message filtering generally relies on the stationarity of the messages  for consistency rather than consistency with consistency in sequential actions expectation. 

In support of our hypothesis,  we see ToM's performance is not impacted by the adaption training. Additionally, we found that the self-interested agent performs better given it has access to shared communication which is to be expected. We have also determined that the performance deterioration was mostly attributed to the false communication messages rather than the presence of an adversary in the environment by allowing the self-interested agent to pass correct messages to the cooperative team, resulting in similar performance to no communication from the self-interested agent. We included the baseline for the self-interested agent with communication (Adv NoDef) because it shows that gains in its performance were direct losses for the cooperative team due to manipulated shared communication.

\begin{figure}
    \centering
    \includegraphics[width=1.0\linewidth]{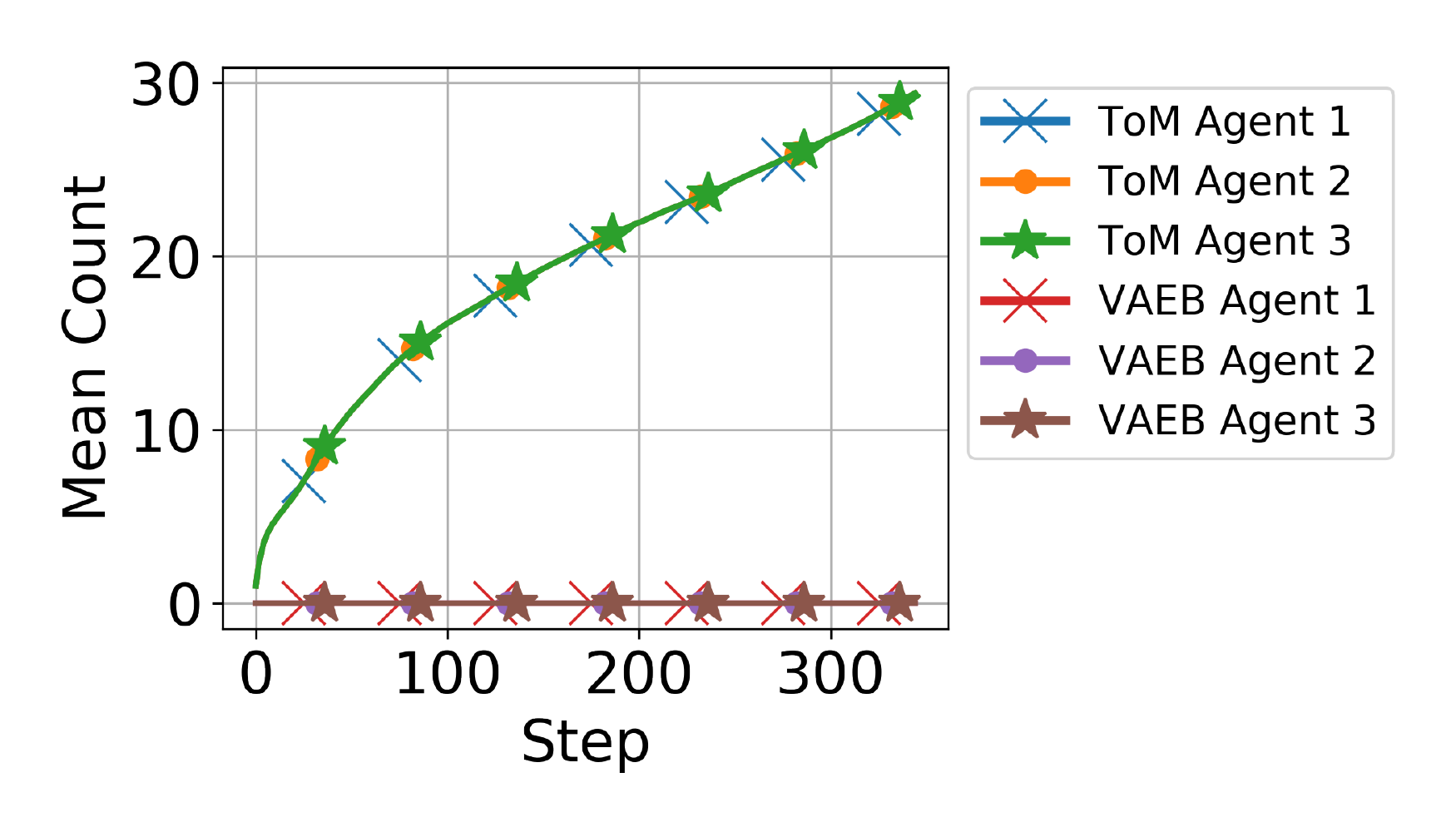}
    \caption{False negative (trusted adversary) agent mean count with readaption training}
    \label{mean_fnc_per_team}
\end{figure}

\begin{figure}
    \centering
    \includegraphics[width=1.0\linewidth]{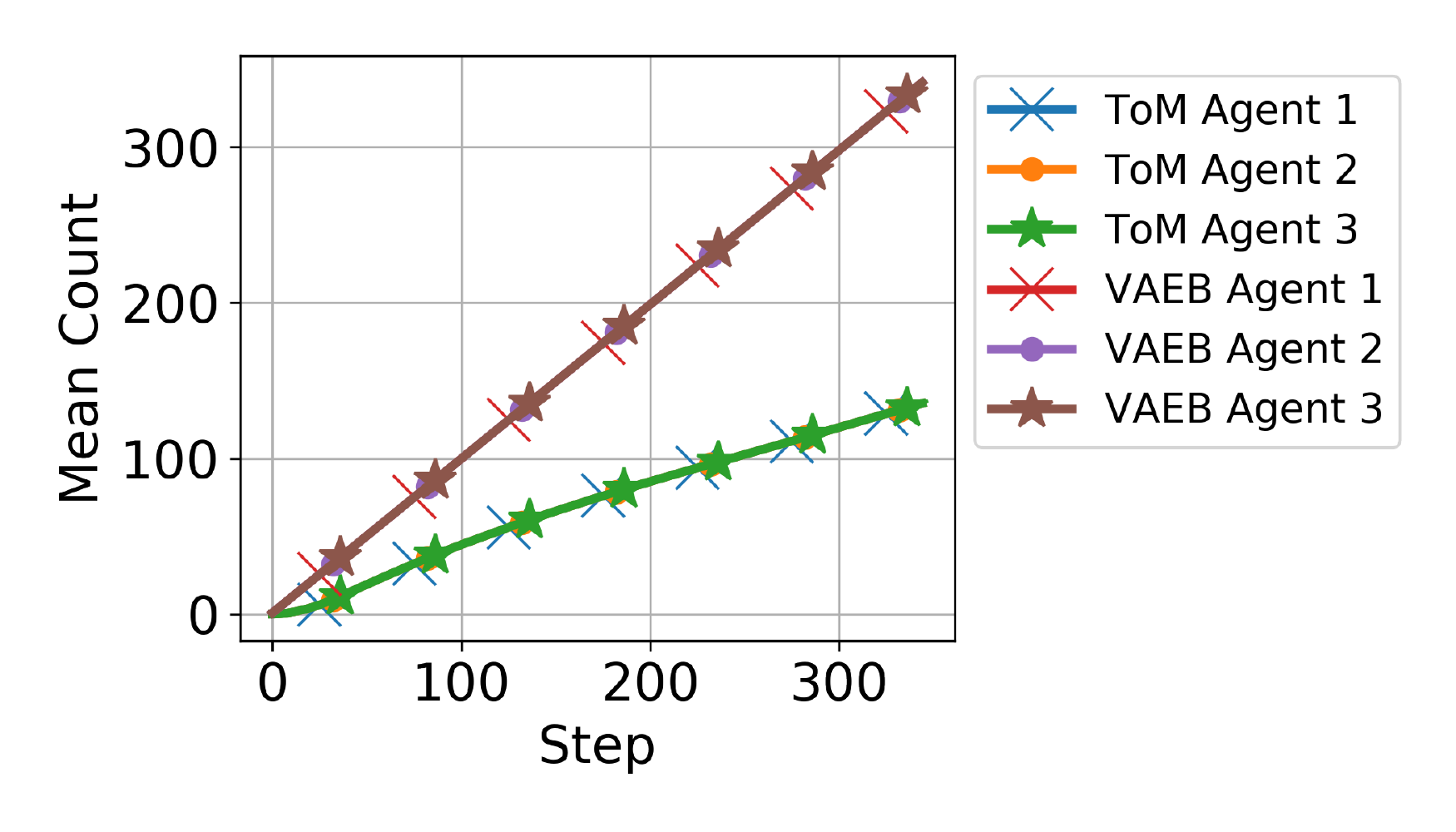}
    \caption{False positive (distrusted cooperative) mean count with readaption training}
    \label{mean_fpc_per_team}
\end{figure}

\begin{figure}
    \centering
    \includegraphics[width=1.0\linewidth]{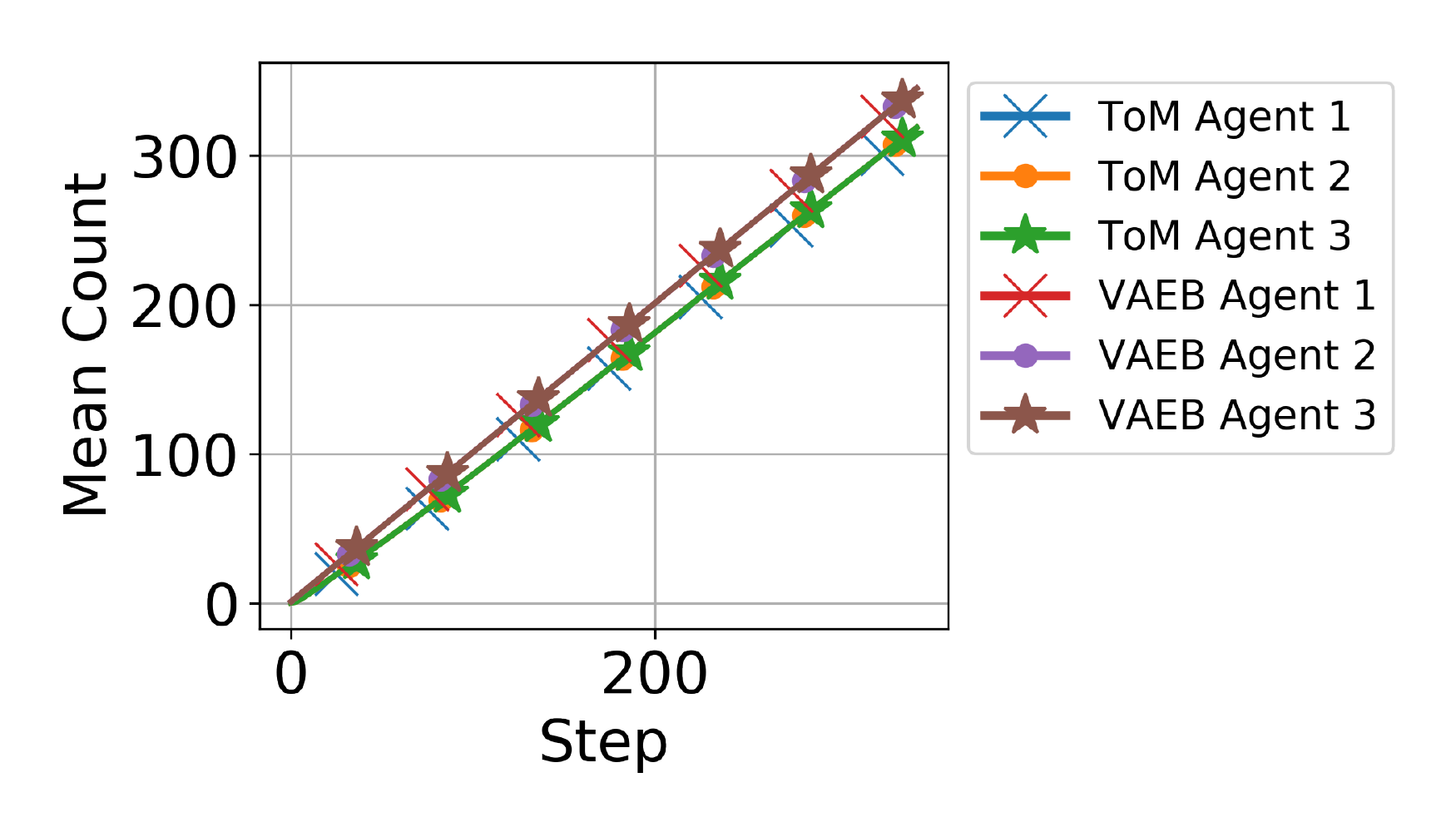}
    \caption{True positive adversary detection mean count with readaption training}
    \label{mean_tpc_per_adv_team}
\end{figure}

\begin{figure}
    \centering
    \includegraphics[width=1.0\linewidth]{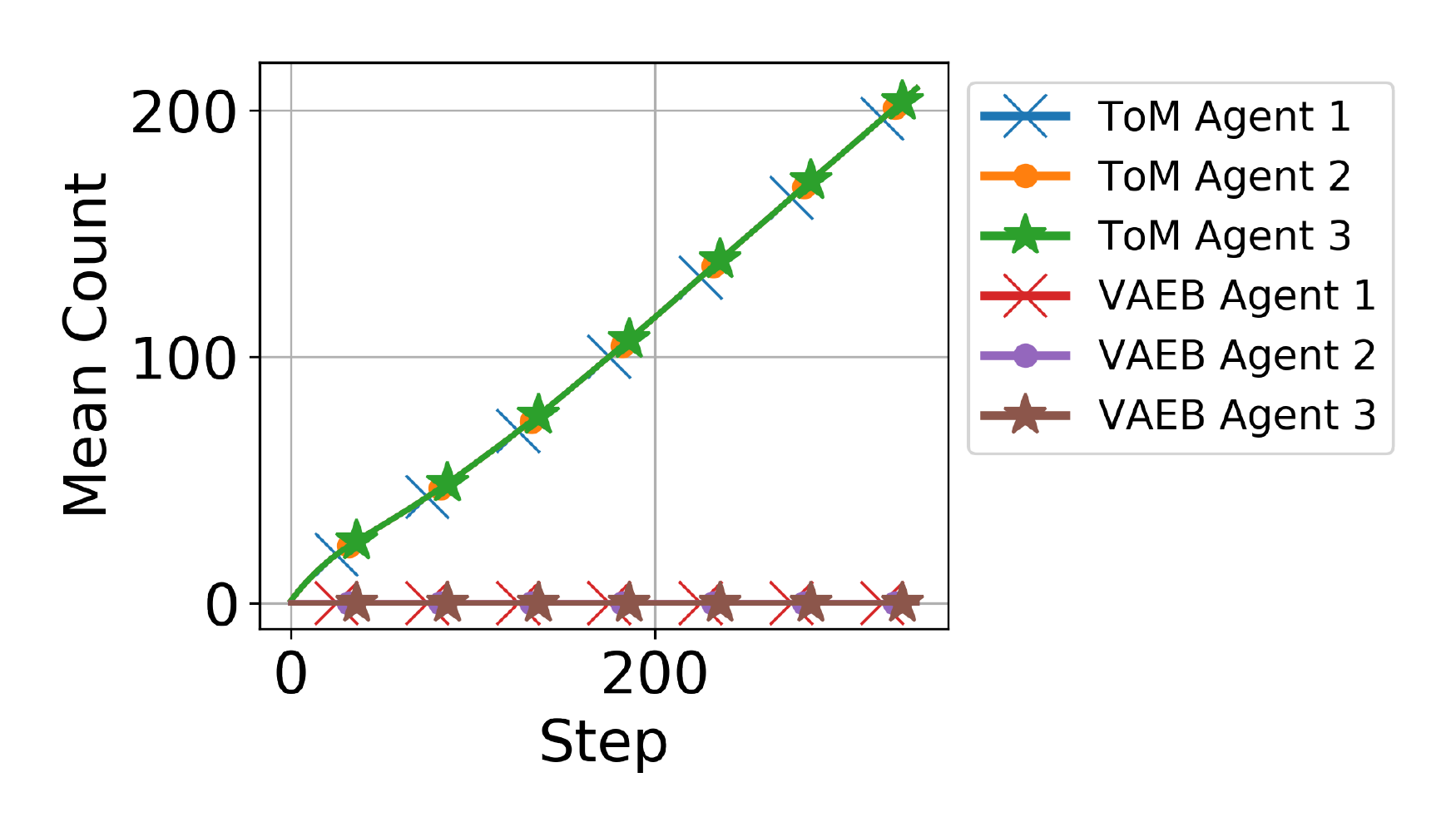}
    \caption{True positive cooperative detection mean count with readaption training}
    \label{mean_tpc_per_team}
\end{figure}

\begin{figure}
    \centering
    \includegraphics[width=1.0\linewidth]{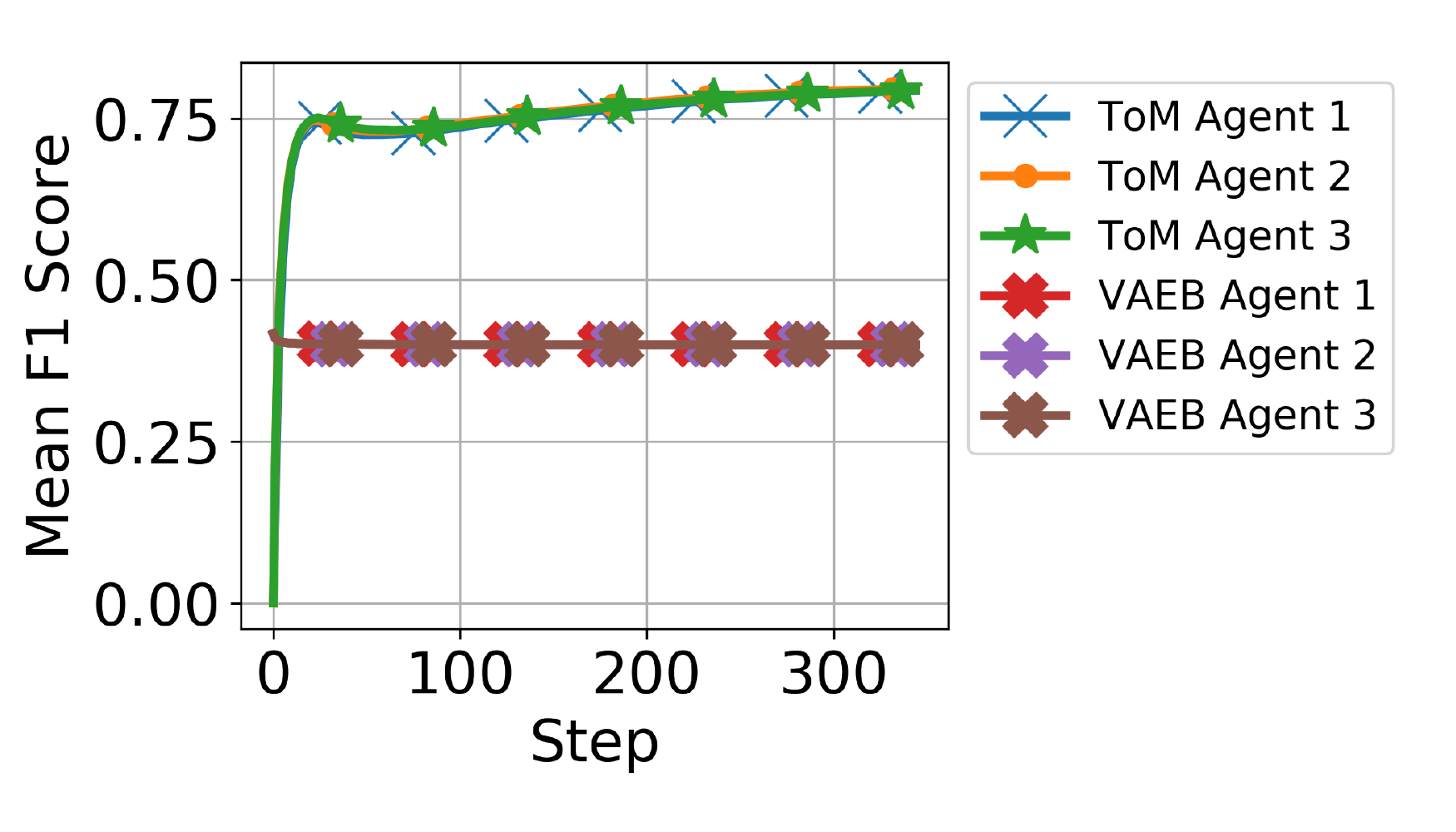}
    \caption{Mean f1 score across steps with readaption training}
    \label{mean_f1}
\end{figure}

We provide Figure \ref{mean_fpc_per_team} and Figure \ref{mean_fnc_per_team} per agent which is the false negative and false positive mean count respectively. We see that since there were more false positives than false negatives as well as Figures \ref{mean_tpc_per_team} and \ref{mean_tpc_per_adv_team} for true positive mean count of cooperative agents and true positive mean count for non-cooperative/self-interested agent, our defense by implicit design was prone to distrusting agents over trusting agents. Additionally, while the VAEB defense is effective at rejecting out-of-distribution messages for a fixed cooperative message distribution, it results in more false-positives when readaption training does not attempt to constrain message learning to be within the current VAEB's message distribution. We also include Figure \ref{mean_f1} which is the f1 score of each agent over steps after adaption training, showing that with more steps, the higher the f1 score for our defense while the VAEB becomes stagnant most likely due to the increase of false positives. Our defense's performance after readaption training did not deteriorate implying this additional defense is both relatively cheaper than consistent retraining and is flexible when layered on top of other defenses.

One other notable direction we tried is to craft adversarial messages; however, we had little success for generating successful targeted attacks with the AGNN architecture rather than the VAEB. Many attempts lead us to believe evasion attacks where we wish to find adversarial samples that can change the actions for all other agents, particularly but not limited to targeted attacks, is likely more difficult than finding adversarial samples that evade a single agent with a standard deep learning architecture.


\section{Conclusion}
In this paper, we have introduced Theory of Mind (ToM) as a rationalization of other agents' behaviors through observed actions to determine trust beliefs of other agents' communicated messages. Our defense is for test-time and requires no additional training or retraining when cooperative agents are allowed to adapt and can be easily layered with other defenses. We study a multi-agent environment where adversarial communication emerges and demonstrate the usage of the ToM trust mechanism defense, comparing the average cooperative team's performance with our defense in comparison to the average cooperative team performance without a defense and cooperative team's performance with a variational auto-encoder bayes defense baseline.

\section{Citations and References}





\bibliographystyle{ACM-Reference-Format} 
\bibliography{sample}


\end{document}